\documentclass[aps,pra,twocolumn,showpacs,superscriptaddress,amsmath,amssymb]{revtex4-1}
\usepackage{mathrsfs}
\usepackage{graphicx}% Include figure files
\usepackage{bm}% bold math
\usepackage{float}
\usepackage{physics}
\usepackage{siunitx}
\usepackage{color}
\usepackage[colorlinks,linkcolor=blue,anchorcolor=blue,urlcolor=blue,citecolor=blue]{hyperref}

\begin{document}

\title{Scheme and Experimental Demonstration of Fully Atomic Weak Value Amplification}

\author{Chun-Wang Wu}
 \thanks{These authors contributed equally to this paper.}
\author{Jie Zhang}
 \thanks{These authors contributed equally to this paper.}
\author{Yi Xie}
\author{Bao-Quan Ou}
\author{Ting Chen}
\author{Wei Wu}
 \email{Corresponding author: weiwu@nudt.edu.cn}
\author{Ping-Xing Chen}
 \email{Corresponding author: pxchen@nudt.edu.cn}
\affiliation{Department of Physics, College of Liberal Arts and Sciences, National University of Defense
 Technology, Changsha 410073, Hunan, China.}
\affiliation{Interdisciplinary Center for Quantum Information, National University of Defense
 Technology, Changsha 410073, Hunan, China.}

\date{\today}

\begin{abstract}
 In this paper, we explore the possibilities of realizing weak value amplification (WVA) using purely atomic degrees of freedom. Our scheme identifies the internal electronic states and external motional states of a single trapped $^{40}$Ca$^+$ ion as the system degree and pointer degree respectively, and their controllable weak coupling is provided by a bichromatic light field. In our experimental demonstration, by performing appropriate postselection on the internal states, a position displacement of $\sim \SI{4}{\angstrom }$ (in phase space) of the trapped ion is amplified to $\sim \SI{10}{\nano\meter }$.
 The sensitivity
 of the amplification effect to the relative phase of the quantum state is also demonstrated. The high operational flexibility of this procedure allows fully exploration of the peculiarities of WVA.
\end{abstract}

\pacs{03.\,65.\,Ta, 39.\,20.\,+q, 06.\,20.\,Dk, 42.\,50.\,Ct}

\maketitle

 Quantum measurement lies at the heart of quantum mechanics, and its peculiar properties have perplexed physicists for almost one
 century \cite{reve1}. In 1932, von Neumann proposed a phenomenological model which describes quantum measurement as an interaction
 process between two quantum systems (or two quantum degrees of one system), regarded as ``measured system"  and ``measuring pointer" respectively \cite{reve2}. Then in 1988, Aharonov, Albert and Vaidman (AAV) extended
 von Neumann's measurement model from the regular strong measurement regime to the weak measurement regime, corresponding
 to the cases where the system and pointer interact strongly and weakly, respectively \cite{reve3}. They showed that, if an appropriate
 postselection step is introduced into the weak measurement procedure, the measured value (defined by AAV as ``weak value") may lie outside
 the eigenvalue spectrum of the observable in question, giving rise to the so-called ``weak value amplification (WVA)" effect. In recent
 years, WVA has been proven to be useful as an amplification technique for observing the extreme weak physical effect, quantum state
 tomography, and tests of quantum mechanics paradoxes \cite{reve4,reve5}.

 In virtue of the high-level manipulation and mature detection techniques for photons, WVA has made great development in the linear optical
 system both in theoretical and experimental aspects. Up to now, optical WVA technique has been successfully applied to the observation of
 the spin Hall effect of light \cite{reve6}, the ultrasensitive measurement of the light beam deflection \cite{reve7}, the amplification of
 optical nonlinearities \cite{reve8}, the direct tomography of photonic quantum states \cite{reve9}, the clarification of controversial debates
 in quantum mechanics \cite{reve10}, and the counteraction of decoherence in the quantum optical system \cite{reve11}. Although WVA cannot
 overcome the shot-noise-limit imposed by quantum mechanics \cite{reve12,reve13}, some references claimed that this amplification
 technique may offer substantial improvements in many practical cases where certain types of technical noise and operational limitations
 dominate \cite{reve14,reve15,reve16,reve17,reve18,reve19}.

 Recently, much effort has been expended in attempts to realize WVA in at least partially nonphotonic systems \cite{reve20,reve21,reve22,reve23,
 reve24,reve25,reve26,reve27}. Among these works, exploration of WVA in cold atoms or trapped ions is particularly interesting. Compared with optical WVA which can be usually
 explained using classical optics \cite{reve28}, atomic WVA is a purely quantum effect and is more suitable to verify the peculiarities of
 quantum properties of WVA. In 2013, Shomroni \emph{et al.} demonstrated WVA based on atomic spontaneous emission \cite{reve25}. However, the involving
 atomic ensemble only plays the role of light source in their procedure and it is still an optical WVA experiment in essence. More recently,
 Ref.\,\cite{reve26} presented a WVA proposal based on atomic matter wave interferometry, but the requirement for accurate manipulation and
 detection of the external orbital motions of free atoms makes it difficult to perform in experiment. Then in 2019, Araneda \emph{et al.}
 performed WVA based on spontaneous emission of a single trapped ion to calibrate the wavelength scale errors in optical
 localization \cite{reve27}. To the best of our knowledge, the experimental demonstration of WVA using purely atomic degrees of freedom has
 not been realized so far.

 In this paper, we present the first proof-of-principle experimental demonstration of purely atomic WVA using a single trapped $^{40}$Ca$^+$ ion.
 For our experiment, the internal electronic states and the external motional states of the ion play the roles of the measured system and the
 measuring pointer, respectively. With the help of spin-dependent displacement, heralded measurement and motional wavepacket reconstruction techniques
 well established in the trapped ion system, we achieve an effective amplification factor as large as $25$ for the position displacement (in phase space) of the ion.
 We also show that WVA is sensitive to the relative phase of the quantum state. We believe that this work opens up the interesting
 possibility to explore WVA in atomic interferometric techniques.

 The WVA formalism can be briefly outlined based on von Neumann's measurement model \cite{reve3}. A quantum system, with its observable $\hat{A}$ being
 measured, is coupled to a measuring pointer by an interaction Hamiltonian $\hat{H}=g\delta(t-t_{0})\hat{A}\hat{p}$, where $\hat{p}$ is the canonical
 momentum of the pointer (with conjugate position $\hat{q}$), $g$ is the interaction strength, $t$ is the time variable, and $\delta(t-t_{0})$ is the impulse function centered at the time ${t_{0}}$ satisfying $\int_{-\infty}^{\infty}\delta(t-t_{0})dt=1$. Suppose the measured system is prepared in the state
 $\ket{\psi_i}=\sum_m\alpha_m\ket{a_m}$, where $\ket{a_m} $ is the eigenstate of $\hat{A}$ with the corresponding eigenvalue $a_m$, and the measuring
 pointer is initialized in $\ket{\varphi(q)}$, where $q$ is the position variable. Under the action of the unitary evolution operator $\hat{U}=\exp(-ig\hat{A}\hat{p}/\hbar)$, the combined
 state will evolve into $\sum_m\alpha_m\ket{a_m}\ket{\varphi(q-a_mg)}$. In the weak coupling limit where $g$ is so small that $\hat{U}$ needs only
 to be expanded to first order in perturbation theory, if we introduce a postselection step by performing a projective measurement on the system and retaining
 only instances with given outcome $\ket{\psi_f}$, the pointer will be left in the state
  \begin{eqnarray}
 \matrixel{ \psi_f}{\hat{U}}{\psi_i}\ket{\varphi(q)}  & \approx &
 \matrixel{ \psi_f}{(1-ig\hat{A}\hat{p}/\hbar)}{\psi_i}\ket{\varphi(q)}\nonumber\\
 & = & \braket{\psi_f}{\psi_i}(1-iA_wg\hat{p}/\hbar)\ket{\varphi(q)}\nonumber\\
 & \approx & \braket{\psi_f}{\psi_i}\exp(-iA_wg\hat{p}/\hbar)\ket{\varphi(q)} \nonumber\\
 & = & \braket{\psi_f}{\psi_i}\ket{\varphi(q-A_wg)},
  \end{eqnarray}
 where
 \begin{equation}
 A_w\equiv\matrixel{\psi_f}{\hat{A}}{\psi_i}/\braket{\psi_f}{\psi_i}
 \end{equation}
 is the weak value defined by AAV. It is easy to note that, under the condition of $\abs{\braket{\psi_f}{\psi_i}}\ll1$, the pointer shift is ``amplified" because $\abs{A_w}\gg\max_m\{\abs{a_m}\}$ and the so-called WVA effect arises. It can be shown from {\color{blue}{Eq.\,(2)}} that the weak value $A_w$ may be an imaginary number. For the imaginary WVA, Ref.\,\cite{reve29} showed that we can observe an amplified shift in the momentum direction with the same interaction Hamiltonian above.

 Our scheme of WVA in a single trapped ion can be simply illustrated in {\color{blue}{Fig.\,1(a)}}. Firstly, a pseudo-spin, composed of two internal electronic states  of the ion and used as the system degree, is prepared in the superposition of the eigenstates of one system observable being measured. The external motion
 of the ion, used as the pointer degree, is initialized to the ground state wavepacket. Then, a weak spin-dependent force displaces the motional wavepacket
 towards opposite directions in phase space for different eigenstates of the system observable. Finally, by performing the projective measurement of another system
 observable and retaining the postselected outcome, the original tiny splitting of the wavepacket may be changed to a large displacement towards a certain
 direction in phase space. This striking amplification effect essentially results from the constructive quantum interference in the tail of the splitted
 wavepackets and destructive interference elsewhere. While introduced as a model of the measurement of a system observable, this procedure is usually
 used as an amplification method for the weak interaction between the system degree and the pointer degree.

 \begin{figure}[!t]
 \includegraphics*[scale=0.85]{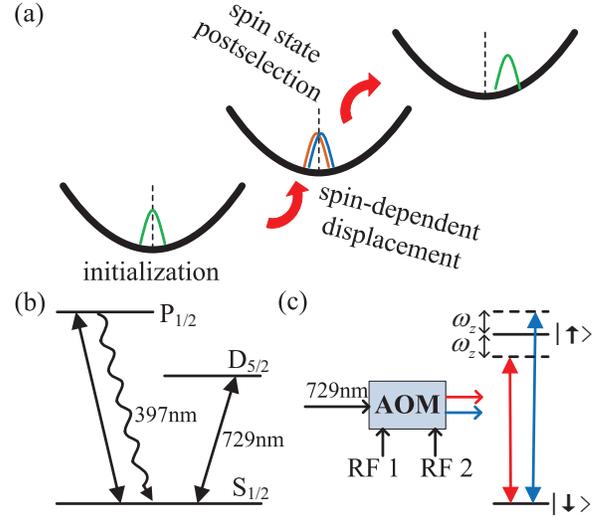}
 \caption{\label{fig1}(Color online) (a) Schematic illustration of WVA effect in a single trapped $^{40}$Ca$^+$ ion. (b) The internal electronic states
 and coupling lasers used for coherent manipulation and projective measurement in our experiment. (c) Configuration of the bichromatic light field used for
 spin-dependent displacement. }
 \end{figure}

 In the experiment, we use a single $^{40}$Ca$^+$ ion trapped in a blade-shaped linear Paul trap with radial and axial trapping frequencies $\omega_r=2\pi
 \times\SI{1.6}{\mega\hertz}$ and $\omega_z=2\pi\times\SI{1.41}{\mega\hertz}$, respectively. {\color{blue}{Fig.\,1(b)}} shows the relevant internal levels and coupling lasers used
 for coherent manipulation and projective measurement (more details on the experimental configuration can be found in the appendix). The Zeeman
 sublevels $S_{1/2}(m_J=-1/2)$ and $D_{5/2}(m_J=-1/2)$ in a magnetic field of $5.3$ G are identified as the spinor states $\ket{\downarrow}$ and
 $\ket{\uparrow}$ respectively, the optical quadrupole transition between which is driven by a narrow linewidth laser at $\SI{729}{\nano\meter}$. The $\SI{729}{\nano\meter}$ laser beam
 enters the trap by passing through two hollow end-cap electrodes with an angle of $0^\circ$ to the axial $z$ direction, resulting in a Lamb-Dicke parameter
 of $\eta\simeq0.08$. The spinor state measurement is done via state-dependent fluorescence observed while coupling the $S_{1/2}$ state to the short-lived
 state $P_{1/2}$ (with lifetime $\SI{7.1}{\nano\second}$) using a laser field at $\SI{397}{\nano\meter}$. The external motional degree we use for the experiment is the axial mode
 of the motion, which can be treated as a quantum mechanical harmonic oscillator with a frequency of $\omega_z$. Its ground-state wavepacket has the size of
 $\Delta_z=\sqrt{\hbar/{2m\omega_z}}\simeq\SI{9.47}{\nano\meter}$, where $m$ is the ion's mass.

 The spin-dependent displacement is implemented using a bichromatic light field resonant with both the blue and red axial sidebands of the $\ket{\downarrow}\leftrightarrow\ket{\uparrow}$ transition, which is realized by sending the \SI{729}{\nano\meter} laser field through an acousto-optic modulator (AOM)
 driven with two frequencies of RF signals that differ by $2\omega_z$ (see {\color{blue}{Fig.\,1(c)}}). In the Lamb-Dicke regime, the resulting Hamiltonian reads
   \begin{eqnarray}
 \hat{H}_d & = & \frac{\hbar\eta\Omega}{2}[\hat{\sigma}_x\sin{\phi_{+}}+\hat{\sigma}_y\cos{\phi_{+}}]  \nonumber\\
 & & \otimes[-(\hat{a}^{\dag}+\hat{a})\cos{\phi_{-}}+i(\hat{a}^{\dag}-\hat{a})\sin{\phi_{-}}],
 \end{eqnarray}
 where $\Omega$ is the Rabi strength, $\hat{a}^{\dag}$ and $\hat{a}$ are the creation and annihilation operators for the axial mode of motion, $2\phi_+=\phi_{red}+\phi_{blue}$ and $2\phi_-=\phi_{red}-\phi_{blue}$ are the sum and the difference of the red sideband laser phase $\phi_{red}$ and the blue
 sideband laser phase $\phi_{blue}$. It can be noted from {\color{blue}{Eq.\,(3)}} that, the setting of $\phi_-$ determines the axis of the displacement in phase space, the internal state of the ion determines the displacement direction along this axis, and $\phi_+$ determines which eigenstates are selected.

 It has to be stressed that {\color{blue}{Eqs.\,(1)}} and {\color{blue}{(2)}} only hold in the weak coupling limit \cite{reve30,reve31,reve32}. In the following, we describe the concrete experimental procedure based on the accurate analytic derivation which is valid for arbitrary coupling strength.

 After Doppler cooling, sideband cooling, and optical pumping, the internal state of the trapped ion is prepared in $\ket{\psi_i}=\ket{\downarrow}$, and its motional state is initialized to the ground state $\ket{\varphi_i(z)}=(\frac{1}{2\pi{\Delta_z}^2})^{\frac{1}{4}}\exp(-\frac{z^2}{4{\Delta_z}^2})$. By setting $\phi_+=\phi_-=\frac{\pi}{2}$, the spin-dependent displacement Hamiltonian has the form $\hat{H}_d=\eta\Omega\Delta_z\hat{\sigma}_x\hat{p}$ with the momentum operator $\hat{p}=\frac{i\hbar(\hat{a}^{\dag}-\hat{a})}{2\Delta_z}$. Application of this Hamiltonian for a duration $t$ generates the entangled state of system and pointer degrees as
 \begin{equation}
 \ket{\Psi}=\frac{1}{\sqrt{2}}[\ket{+}\ket{\varphi_i(z-g\Delta_z)}-\ket{-}\ket{\varphi_i(z+g\Delta_z)}],
 \end{equation}
 where $\ket{+}=\frac{1}{\sqrt{2}}(\ket{\uparrow}+\ket{\downarrow})$, $\ket{-}=\frac{1}{\sqrt{2}}(\ket{\uparrow}-\ket{\downarrow})$ and $g=\eta\Omega t$. For the trapped ion system, we can only postselect the state $\ket{\uparrow}$ without destroying the motional wavepacket as no photons are scattered in the measurement process \cite{reve33,reve34}. However, the postselections of other states can be carried out with the help of appropriate one-qubit rotations. In the experiment, we firstly perform a rotation by $2\theta$ around the y-axis of the Bloch sphere (denoted as $R_y(2\theta)$) on the internal state, which is realized using a laser pulse resonant with the $\ket{\downarrow}\rightarrow\ket{\uparrow}$ transition, then perform the strong projective measurement via state-dependent fluorescence. After retaining only instances with measurement outcome $\ket{\uparrow}$, we effectively realize the postselection with $\ket{\psi_f}=R_y(-2\theta)\ket{\uparrow}=\cos\theta\ket{\uparrow}-\sin\theta\ket{\downarrow}$. Then, the external motion will be left in the state (choosing $\Delta_z$ as the unit of $\hat{z}$)
 \begin{eqnarray}
 \ket{\varphi_f(z)} & = & \frac{1}{(2\pi)^\frac{1}{4}\sqrt{1-\cos(2\theta)e^{-\frac{g^2}{2}}}}[\cos(\frac{\pi}{4}+\theta)e^{-\frac{(z-g)^2}{4}}  \nonumber\\
 && -\sin(\frac{\pi}{4}+\theta)e^{-\frac{(z+g)^2}{4}}],
 \end{eqnarray}
 with the position displacement in phase space
 \begin{equation}
 \delta_z=\matrixel{\varphi_f(z)}{\hat{z}}{\varphi_f(z)}=\frac{g}{e^{-\frac{g^2}{2}}\cot(2\theta)-\csc(2\theta)}.
 \end{equation}
 It follows from {\color{blue}{Eqs.\,(1)}} and {\color{blue}{(2)}} that, the position displacement will reduce to $\delta_z\approx\frac{\matrixel{\psi_f}{\hat{\sigma}_x}{\psi_i}}{\braket{\psi_f}{\psi_i}}g=-\cot(\theta)g$ in the weak coupling limit.

 The only observable of the trapped ion that can be directly measured via state-dependent fluorescence is $\hat{\sigma}_z$. To proceed with the analysis of the postselected motional state, we use an indirect measurement method for motional wavepacket reconstruction, which was originally proposed in 1995 \cite{reve35} and has been successfully utilized in the quantum simulation of Dirac equation \cite{reve36} and random walk \cite{reve37} in trapped ion system. To measure the marginal probability distribution of the motional state along a line in phase space, we prepare the internal state of the ion in an eigenstate of $\hat{\sigma}_y$ or $\hat{\sigma}_z$, apply a spin-dependent displacement operation that displaces the motional wave function in phase space in a direction orthogonal to the one to be measured for varying amounts of time, followed by a measurement of the changing excitation of the ionic internal state. Then, a Fourier transformation of these measurements yields the marginal probability distribution of the motional wavepacket. Furthermore, the expectation values of the motional quadratures can be obtained by analyzing the slope of the changing excitation in very short time, without a need to reconstruct the full motional wavepacket (see Appendix for details).

\begin{figure}[!t]
	\includegraphics*[scale=0.445]{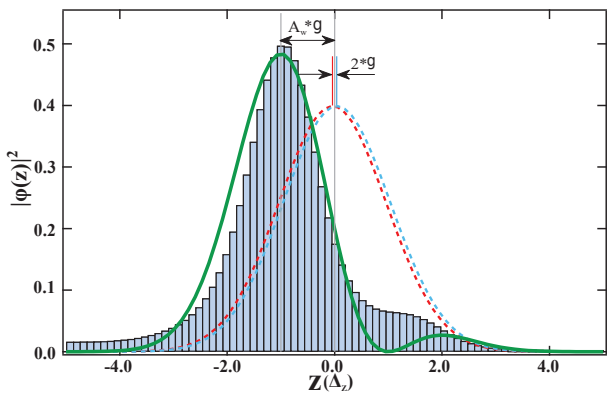}
	\caption{\label{fig2}(Color online) A position displacement of $\SI{4}{\angstrom}$ (in phase space) of the trapped ion is amplified to $\SI{10}{\nano\meter}$. The dashed lines indicate
    the probability distributions of the two weakly displaced wavepackets correlated with the eigenstates of $\hat{\sigma}_x$ in {\color{blue}{Eq.\,(4)}} with $g\simeq0.04$. The histogram shows the probability distribution of the finally postselected wavepacket reconstructed using the experimental data with $\theta=0.02$ and the solid line gives the exact theoretical prediction using {\color{blue}{Eq.\,(5)}}. For this parameter regime with the low success rate of postseletion, both the spin state detection error and the instability of experimental parameters during the long time scale of data acquisition result in the difference between the measured distribution and the theoretical prediction in {\color{blue}{Fig.\,2}}.}	
 \end{figure}

 \begin{figure}[!t]
	\includegraphics*[scale=0.52]{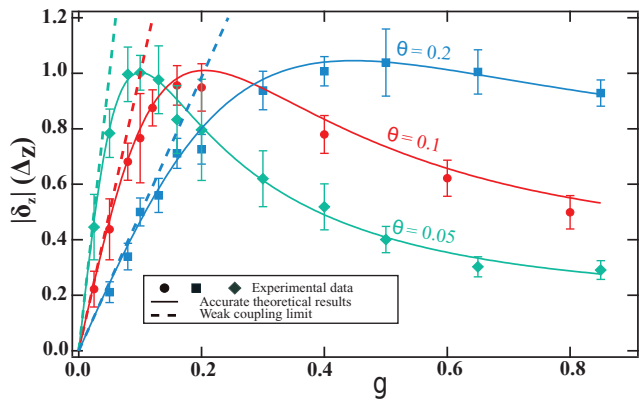}
	\caption{\label{fig3}(Color online) The relation between the magnitude of the amplified position shift in phase space $\abs{\delta_z}$ and the original splitting $g$ with
    different $\theta$. The dashed lines represent the theoretical results in the weak coupling limit where we have $\abs{\delta_z}\approx g\abs{A_w}=g\cot{\theta}$. The solid lines are the accurate theoretical results. The symbols ($\bullet$, $\blacksquare$ and $\blacklozenge$) are the experimental data. The error bars represent the fitting errors using the weighted fitting method based on the population data with quantum projection noise(see Appendix).}
 \end{figure}

 \begin{figure*}[ht]
	\centering
	\includegraphics*[scale=0.5]{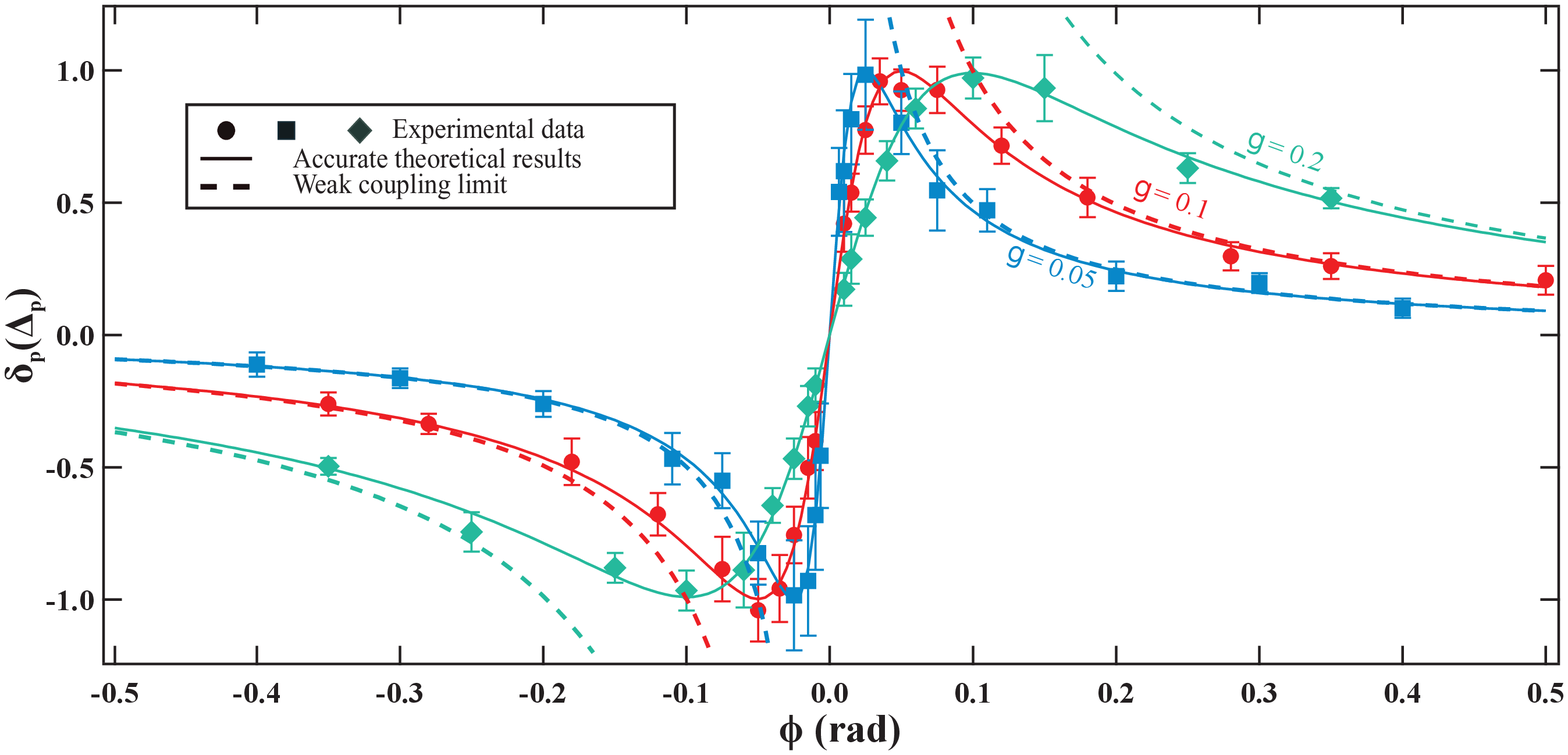}
	\caption{\label{fig4}(Color online) By choosing $\ket{\psi_i}=R_x(2\phi)\ket{\downarrow}$ and $\ket{\psi_f}=\ket{\uparrow}$, we achieve the pure
    imaginary weak value $A_w=i\cot\phi$ and herein the original small position splitting $g$ will result in an amplified momentum shift $\delta_p$ in phase space. The dashed lines represent the theoretical results in the weak coupling limit where we have $\delta_p\approx g\Im(A_w)=g\cot{\phi}$. The solid lines are the accurate theoretical results. The symbols ($\bullet$, $\blacksquare$ and $\blacklozenge$) are the experimental data. The error bars represent the fitting errors using the weighted fitting method based on the population data with quantum projection noise(see Appendix).}
 \end{figure*}

 In {\color{blue}{Fig.\,2}}, we experimentally show that a position displacement of $\SI{4}{\angstrom}$ (in phase space) of the trapped ion can be amplified to $\SI{10}{\nano\meter}$ using the above WVA procedure. In the experiment we use a short spin-dependent displacement pulse with $\Omega=2\pi\times \SI{19.0}{\kilo\hertz}$ and $t=\SI{4 }{\micro\second}$ to achieve a small
 position splitting $g\simeq0.04$. After the internal state postselection using $R_y(-2\theta)\ket{\uparrow}$ with $\theta=0.02$, a remarkable WVA effect occurs with the amplified position displacement $\SI{10}{\nano\meter}$. It should be noted that the weak coupling limit does not hold for this parameter setting and the theoretical amplified position displacement in {\color{blue}{Fig.\,2}} is actually obtained from the exact result of {\color{blue}{Eq.\,(6)}}. The current limitation in the amplification factor is given by the low success rate of postselection which requires stability of the experimental parameters during the long time scale of data acquisition.

 Through fast measurement of the expectation value of the position operator, we show the relation between the amplified position shift $\abs{\delta_z}$ in phase space and the original splitting $g$ with different $\theta$ in {\color{blue}{Fig.\,3}}. In the weak coupling regime defined by $\abs{A_w}g\ll1$ \cite{reve38}, $\abs{\delta_z}$ increases linearly with $g$ with an amplification factor $\abs{A_w}=\cot\theta$ and the approximate results agree well with the accurate theoretical results and the experimental data. Outside this regime, the amplified position shift has an upper limit $\sim\Delta_z$. With $g$ increasing, the weak measurement regime will transit to the strong measurement regime and the WVA effect disappears.

 The WVA procedure with the pure imaginary weak value can also be implemented in our experiment. By choosing $\ket{\psi_i}=R_x(2\phi)\ket{\downarrow}=\frac{1}{\sqrt{2}}(\ket{+}-e^{i2\phi}\ket{-})$ and $\ket{\psi_f}=\ket{\uparrow}$, we achieve the pure imaginary weak value $A_w=i\cot\phi$. In this case, the original small position splitting $g$ results in an amplified moment shift (choosing $\Delta_p=\frac{\hbar}{2\Delta_z}$ as the unit of $\hat{p}$)
 \begin{equation}
 \delta_p=\frac{g\sin(2\phi)}{e^{\frac{g^2}{2}}-\cos(2\phi)}
 \end{equation}
 in phase space, as shown in {\color{blue}{Fig.\,4}}. In the parameter regime $\abs{A_w}g\ll1$, the weak coupling approximation is valid and its predictions agree well with the accurate theoretical results and the experimental data. Outside this regime, the amplified momentum shift has an upper limit $\sim\Delta_p$.

 We have demonstrated the realization of WVA using purely atomic degrees of freedom. An effective amplification factor of $25$ for the position displacement of the ion is achieved, which is much smaller than optical WVA (e.g. an amplification factor bigger than $127$ is obtained in Ref.\,\cite{reve7}). The current limitation in the amplification factor of this procedure is the long time scale of data acquisition which requires stability of the experimental parameters. The lack of direct measurement method for motional state is another challange for studying WVA using trapped ion system. Considering these existing limitations, whether this procedure can afford benefits in metrological applications is still an open question. However, our work has demonstrated the high operational flexibility of atomic WVA , which makes it an important complement to optical WVA. The interaction strength between the measured system degree and the measuring pointer degree can be tuned over a wide range, and their coupling pattern can be manipulated by adjusting the experimental parameters. Combined with the easily tailored pointer state, which can be prepared in various nonclassical motional states (e.\,g.\,Fock states, squeezed states, even two-mode NOON states \cite{reve39,reve40}), and the ability to coherently manipulate several tens of ions \cite{reve41}, our approach allows fully exploration of the peculiarities of WVA. The WVA using nonclassical pointer states is particularly worth researching because its potential to improve the measurement precision when applying WVA to metrological scenarios \cite{reve42}.

 In conclusion, we present the first experimental realization of fully atomic WVA using the trapped ion system. The amplification of motional displacement in phase space and the sensitivity of WVA to the relative phase of the quantum state are demonstrated in our experiment. The good agreement between the calculated and experimental results shows the trapped ion as a favorable platform for WVA research. The good scalability and flexible controllability of trapped ion system make it an important complement to optics in this research area. More importantly, our experiment makes the first step towards the study of atomic WVA and has much potential in the studies of fundamental quantum mechanics.

 This work was supported by the National Basic Research Program of China under Grant No.\,2016YFA0301903, the National Natural Science Foundation of China under Grant Nos.\,11174370, 11304387, 61632021, 11305262, and 11574398, and the Research Plan Project of National University of Defense Technology under Grant No.\,ZK16-0304.

\appendix*
\section{Detailed Description of The Experimental Technichs}

In the appendix, we describe the trapped ion system in more detail, discuss the postselection step and the calibration of the experimental parameters, and finally illustrate the wavepacket reconstruction method along with the data analysis for the measurement of $\expval{\hat{z}}$ and $\expval{\hat{p}}$.

\subsection{Ion trap system}

 A single  ion $^{40}$Ca$^+$  trapped in a blade-shaped linear Paul trap is used in the  experiment. In the first part of the experimental sequence, all the motional modes of the single ion are cooled down to a phonon number of about 20 using Doppler cooling method with lasers at 397 and 866 nm. The oscillator we use for our experiment is the axial mode, which has a secular frequency of  around $\omega_z/(2\pi)=1.41$ MHz. After Doppler cooling, the axial mode is cooled further to the motional ground state with the resolved sideband cooling approach \cite{reve43}. A narrow linewidth laser at 729 nm  is used to coherently couple the qubit which we choose the energy levels  $S_{1/2}(m_J=-1/2) $ and  $D_{5/2}(m_J=-1/2) $ as  $\ket{\downarrow}$  and $\ket{\uparrow} $ respectively. The whole setup for controlling the qubit is shown in {\color{blue}{Fig.\,5}}. The power of 729 nm laser beam is stabilized by using acousto-optical modulator (AOM) 1.  AOM 2  is used to shift up  the laser frequency by \SI{80}{\mega\hertz} and modulate the bichromatic light field for the spin-dependent force. The amplitude and  frequency  of the beam are controlled by AOM 3, which has a double pass optical configuration and is driven by an RF source with frequency \SI{270}{\mega\hertz}. It is also used as a switch for generation of laser pulses in the experimental sequence.  The qubit transition is isolated by a gap of 8.8 MHz from its nearest internal state transitions with a magnetic field of 5.3 G. The beam goes through the two end-caps with almost 0 degree with respect to z axis of the trap resulting in a Lamb-Dicke parameter of $\eta\approx0.08$ as shown in {\color{blue}{Fig.\,5}}. Optical pumping to  $\ket{\downarrow}$ is realized using a combination of left-handed circularly polarized  light at 397 nm, and linearly polarized light at 866 nm and 854 nm. The internal state of the ion is read by using the electron shelving technique with a detection time of $\SI{300}{\micro\second}$. The heating rate of the axial motional mode is about 70 quanta per second, and the coherence time of the Fock state superposition $(\ket{0} +\ket{1})/\sqrt{2}$ has been measured to be about  $\SI{5.0}{\milli\second}$. Due to the large magnetic fluctuations induced by the AC-power line, we trigger the experimental cycles  at \SI{50}{\hertz} and the coherence time of  the qubit has been measured using Rasmey fringe to be around \SI{1.1}{\milli\second}.

\begin{figure}
	\centering
	\includegraphics[width=0.9\linewidth]{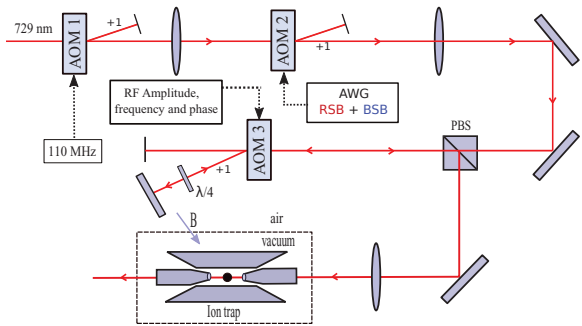}
	\caption{(Color online) The trap geometry and optical setup in our experiment. AOM 1 is used for power stabilization of the 729 nm laser beam. AOM 2
		creates the required bichromatic laser field when it is supplied with two frequencies of RF signals that differ by $2\omega_z$. The RF signals are generated by an arbitrary waveform generator (AWG). AOM 3 controls the overall frequency and amplitude of the 729 nm laser beam. }
\end{figure}

\subsection{HERALDED DETECTION}

The internal state detection via application of lasers at 397 nm and 866 nm is used for postselection of the motional wavepacket. The wavepacket is retained if the ion is in $\ket{\uparrow}$ and no fluorescence photons should be detected, while it is destroyed if the ion is in $\ket{\downarrow}$ and a certain amount of photons should be gathered. Since the coherence time is limited, we want to use the shortest detection time for the postselection to avoid decoherence. The detection time in our experiment is chosen as $\SI{120}{\micro\second}$ and we distinguish the ion's electronic states based on a pre-determined threshold for the gathered photons. With a detection time of $\SI{120}{\micro\second}$ , the detection error is less than 0.003$\%$ for an ion which is measured to be in $\ket{\uparrow}$. As stated in Ref.\,\cite{reve33}, such low error is of the similar size to other sources of error in the experiment.

\subsection{CALIBRATION OF THE DISPLACED OPERATION}
 Making sure that we apply an accurate displacement on the motional ground state is the critical step in our experiment. The displacement operation can be realized by a single bichromatic laser pulse which  creates a Schr\"{o}dinger cat state between the electronic states and the motional states. By applying the spin-dependent displacement operator for a time t on the initial state $\ket{\downarrow}\ket{ 0 }$, where $\ket{ 0 }$ is the motional ground state, we get the entangled state
\begin{equation}\label{1}
\ket{\Psi}=D(\alpha) \ket{\downarrow}\ket{ 0 }=\dfrac{1}{\sqrt{2}}(\ket{+}\ket{\alpha}-\ket{-}\ket{-\alpha}),
\end{equation}
where $\ket{+}=\dfrac{1}{\sqrt{2}}(\ket{\uparrow} +\ket{\downarrow})$, $\mid-\rangle=\dfrac{1}{\sqrt{2}}(\ket{ \uparrow}-\ket{\downarrow})$, and $D(\alpha)= e^\frac{2\alpha\Delta_z\hat{\sigma}_x\hat{p}}{i\hbar}$. The displacement  $\alpha=\eta\Omega t/2$ is proportional to the Lamb-Dicke parameter $\eta$, the coupling strength $\Omega$, and the bichormatic pulse duration t. For the trapped ion, the only observable we can measure is $\hat{\sigma}_z$, and a projective measurement on $ \ket{\Psi}$ returns the probability of ion in $ \ket{\uparrow}$
\begin{equation}\label{2}
p_{\uparrow}=\dfrac{1}{2}(1+\expval{\hat{\sigma}_z})=\dfrac{1}{2}(1-e^{-2\abs{\alpha}^2}).
\end{equation}
Note that  $\Omega$ means both the coupling strengths of red and blue sidebands, hence the intensities of the two RF signals driving AOM 2 are required to be equal. We perform the uniformity calibration of coupling strengths for the red sideband and blue sideband lasers by checking if the probability of $\ket{\uparrow}$ reaches 0.5 when the ground state wavepacket totally  splits into two parts. {\color{blue}{Fig.\,6}} shows  an experimental measurement of
$p_{\uparrow}$ after evolution under the balanced bichromatic field along with the theoretical simulation of {\color{blue}{Eq.\,(\ref{2})}} for $\eta =0.08$ and   $\Omega = 2\pi \times  \SI{150}{\kilo \hertz}$.

\begin{figure}
	\centering
	\includegraphics[width=1.0\linewidth]{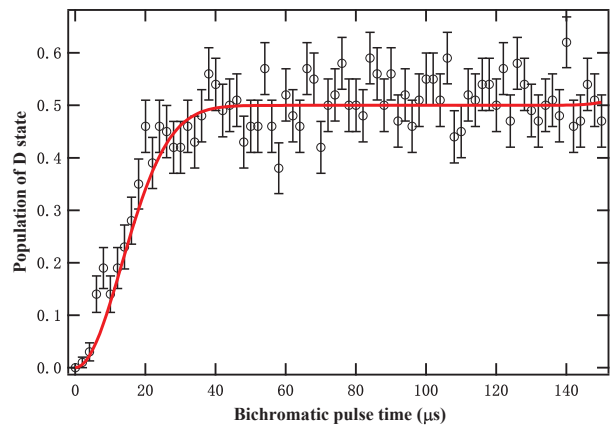}
	\caption{(Color online) Uniformity calibration of coupling strengths for the red sideband and blue sideband lasers. The red line is a plot of
		{\color{blue}{Eq.\,(\ref{2})}} with parameters $\eta =0.08$ and $\Omega = 2\pi \times \SI{150}{\kilo \hertz}$. The black dots indicate the experimental measurement of $\ket{\uparrow}$ at chosen times. }
\end{figure}

To demonstrate WVA, We also need to calibrate the small displacement $\alpha$ as a function of  the bichromatic pulse length by measuring the phonon number of the ion, which follows the Poisson distribution and has the average $\bar{n}=\abs{\alpha}^2$. Experimentally we choose a low coupling strength for the bichromatic light field and probe the average phonon number of the ion by applying the pulse for varying amounts of time, and the corresponding results are shown in  {\color{blue}{Fig.\,7}}. The coupling strength used in the experiment is then determined to be $\Omega\approx 2\pi\times \SI{19.0}{\kilo\hertz}$.

\begin{figure}
	\centering
	\includegraphics[width=1.0\linewidth]{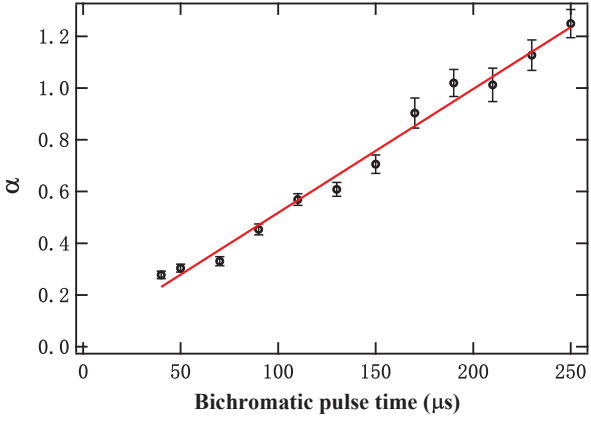}
	\caption{(Color online) Calibration of the displacement $\alpha$ as a function of the bichromatic pulse time with the low coupling strength. The black dots
		indicate $\alpha$ obtained from measurement of the average phonon number $\bar{n}$. The red line denotes the fitting of data points and the corresponding coupling strength  of the bichromatic field is determined to be $\Omega\approx 2\pi\times  \SI{19.0}{\kilo\hertz}$. }
\end{figure}

\subsection{WAVEPACKET RECONSTRUCTION AND DISPLACEMENT MEASUREMENT IN PHASE SPACE}

To obtain the wavepacket probability distribution,  we apply a state-dependent displacement operation $ \hat{U}_z=\exp(-ik\hat{z} \hat{\sigma}_x/2) $ to the measured quantum state and then the following measurement of $\hat{\sigma}_z$ effectively returns the measurement outcome of the observable
\begin{equation}\label{3}
\hat{O}(k)=\hat{U}_z^+\hat{\sigma}_z\hat{U}_z=\cos(k\hat{z})\hat{\sigma}_z+\sin(k\hat{z})\hat{\sigma}_y,
\end{equation}
where $k=\eta\Omega t/\Delta_z$, and  $\Omega \approx 2\pi\times\SI{70}{\kilo\hertz}$. If we prepare the internal state of the ion to be the eigenstate of $\hat{\sigma}_z$ with eigenvalue +1, then we have $ \expval{\hat{O}(k)}= \expval{\cos(k\hat{z})} $. Similarly, we have $\expval{\hat{O}(k)}=\expval{\sin(k\hat{z}) }$ if the eigenstate of $\hat{\sigma}_y$ with eigenvalue +1 is prepared. Theoretically we can get the probability density $\abs{\varphi(z)}^2$ by using Fourier transformation of $\expval{\cos(k\hat{z})}+i\expval{\sin(k\hat{z})}$ \cite{reve36}, however the finite experimental cycles can not afford enough information for this method. Instead we use a constrained least-square optimization method based on convex optimization \cite{reve44} as stated in Ref.\,\cite{reve37}. In this method  the position space is  first discretized by a suitable set of points $z_i$ and  probability distributions of $p(z_i)$ can be searched by minimizing the function
\begin{equation}
\begin{aligned}\label{4}
F=&\sum_k \Big(\sum_i p(z_i)\cos(kz_i)-C_k \Big)^2 \\
&+\sum_k\Big(\sum_i p(z_i)\sin(kz_i)-S_k\Big)^2,
\end{aligned}
\end{equation}
where $C_k$ and $S_k$ are the experimental results of $\expval{ \cos(k\hat{z})}$
and  $\expval{ \sin(k\hat{z})}$ respectively.

In this technique we also need to apply additional constraints. Firstly, the probability of the positon points meet the conditions $0\leq p(z_i)\leq 1$ and $\sum_i p(z_i)=1$. Secondly, $p'(z)$, the differentiation of the probability distribution function $p(z)$ with respect to position $z$, should be bounded by the kinetic energy of the ion \cite{reve45}
\begin{equation}\label{5}
\dfrac{\hbar^2}{8m}\int_{-\infty}^{\infty}dz\dfrac{p'(z)^2}{p(z)} \leq \expval{ \dfrac{\hat{p}^2}{2m}}.
\end{equation}
For obtainment of the value of $\langle \hat{p}^2\rangle$, we only need to adjust the phase of the bichromatic laser pulse to realize $\hat{U}_p=\exp(-ik\hat{p} \hat{\sigma}_x/2) $,
and finally calculate $\langle \hat{p}^2\rangle=d^2/dk^2 \expval{\hat{ O }(k) }$.

\begin{figure}
	\centering
	\includegraphics[width=1\linewidth]{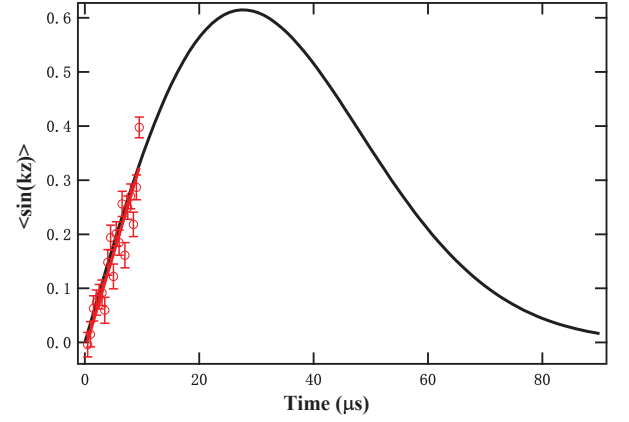}
	\caption{(Color online) Data analysis for obtaining the average displacement. The black curve denotes the full simulation of $\langle \sin(k\hat{z})
		\rangle$ with parameters $g= 0.2$ and $\theta=0.2$. The red cycles are experimental data. The red line shows the weighted fitting of data points for extracting  $\expval{\hat{z}}$.  }
\end{figure}

Besides the wavepacket, we are also concerned about the average shifts of the wavepacket.  From {\color{blue}{Eq.\,(\ref{3}})} we can obtain $\dfrac{d}{dk}\expval{\hat{ O }(k)}\mid_{t=0}=\expval{\hat{z}\hat{\sigma}_y}$, therefore the average position displacement $\expval{ \hat{z} }$ can be measured by preparing the internal state to be the eigenstate of $\hat{\sigma}_y$. Experimentally the state of the ion is $\ket{ \uparrow }$ after postselection, we need to apply  a $\pi/2$ carrier transition laser pulse to coherently prepare the internal state in $(\ket{\downarrow} + i\ket{\uparrow})/\sqrt{2} $. Afterwards we apply the operation $ \hat{U}_z $ and detect $\expval{\sin(k\hat{z})}$ for a short length of time (\SI{10}{\micro\second} in the experiment),  and $\expval{\hat{z}}$ can be  extracted by fitting the data points with a linear model as shown in {\color{blue}{Fig.\,8}}. The same method can be used for the measurement of $\expval{\hat{p}}$, the only difference here is the operation we apply after postselection detection  is  $\hat{U}_p=\exp(-ik\hat{p} \hat{\sigma}_x/2) $.

Since the wavepacket reconstruction pulses are applied after postselection, the success rates of postselection  are  different from point to point in the experiment and it is necessary to consider the different contributions to the fitting error for different data points . Hence, we use the weighted fitting method instead of the regular fitting method to fit the data, the standard deviation derived from quantum projection noise of each data point is used as the weighting parameter \cite{reve45}.  {\color{blue}{Fig.\,8}} shows an example of data analysis for measuring $\expval{\hat{z}}$ with parameters $g=0.4$ and $\theta = 0.2$.

\end{document}